\documentclass[aip,twocolumn,floatfix,reprint]{revtex4-1}

\draft 

\usepackage{graphicx}
\usepackage{dcolumn}
\usepackage{bm}
\usepackage{float}
\usepackage{color}
\usepackage{amsmath}
\usepackage{mathtools}
\usepackage{textcomp}
\usepackage[caption = false]{subfig}
\usepackage{xcolor,colortbl}
\usepackage{xifthen}
\usepackage{MnSymbol}
\usepackage{natbib}
\usepackage{csvsimple}
\usepackage{makecell}
\usepackage{enumitem}
\usepackage[super]{nth}
\usepackage{scalerel}
\usepackage{array}

\begin{document}

\title{Dynamics of supercooled liquids from static averaged quantities using machine learning}

\author{Simone Ciarella}
\email{simoneciarella@gmail.com}
\affiliation{Laboratoire de Physique de l’Ecole Normale Supérieure, ENS, Université PSL, CNRS, Sorbonne Université, Université de Paris, F-75005 Paris, France}
\affiliation{%
	 Soft Matter and Biological Physics, Department of Applied Physics, Eindhoven University of Technology, Den Dolech 2, 5600 MB Eindhoven, The Netherlands
}%

\author{Massimiliano Chiappini}
\affiliation{%
Leonard S. Ornstein Laboratory, 
Princetonplein 1, 3584 CC Utrecht, The Netherlands
}%

\author{Emanuele Boattini}
 \affiliation{%
Leonard S. Ornstein Laboratory, 
Princetonplein 1, 3584 CC Utrecht, The Netherlands
}%
\author{Marjolein Dijkstra}
 \affiliation{%
Leonard S. Ornstein Laboratory, 
Princetonplein 1, 3584 CC Utrecht, The Netherlands
}%
\author{Liesbeth M. C. Janssen}%
 \email{L.M.C.Janssen@tue.nl}
 \affiliation{%
	 Soft Matter and Biological Physics, Department of Applied Physics, Eindhoven University of Technology, Den Dolech 2, 5600 MB Eindhoven, The Netherlands
}%

\date{\today}

\begin{abstract}
We introduce a machine-learning approach to predict the complex non-Markovian dynamics of supercooled liquids from static averaged quantities. 
Compared to techniques based on particle propensity, our method is built upon a theoretical framework that uses as input and output system-averaged quantities, thus being easier to apply in an experimental context where particle resolved information is not available. 
In this work, we train a deep neural network to predict the self intermediate scattering function of binary mixtures using their static structure factor as input. 
While its performance is excellent for the temperature range of the training data, the model also retains some transferability in making decent predictions at temperatures lower than the ones it was trained for, or when we use it for similar systems. 
We also develop an evolutionary strategy that is able to construct a realistic memory function underlying the observed non-Markovian dynamics.
This method lets us conclude that the memory function of supercooled liquids can be effectively parameterized as the sum of two stretched exponentials, which physically corresponds to two dominant relaxation modes.
\end{abstract}


\maketitle 

\section{Introduction}
\begin{figure*}
\centering
\includegraphics[width=\textwidth]{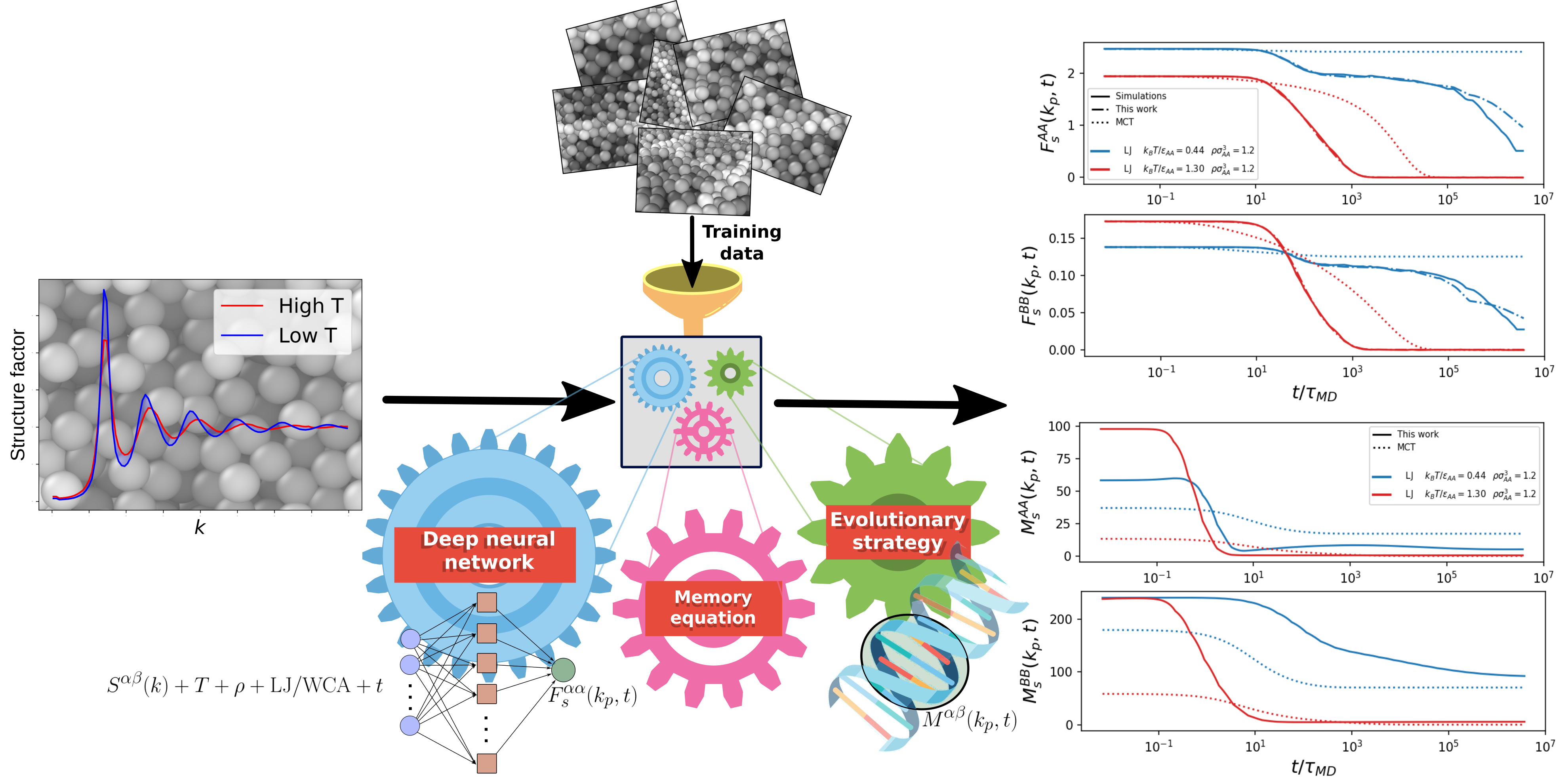}
\caption{Sketch of our machine learning approach based on averaged static descriptors. The static structure factor is the main input of the model and even though it only shows minute changes with the temperature $T$, the model was trained to predict the abrupt slowdown from these minimal changes. The DNN explained in sec.~\ref{sec:dnn} predicts $F^{\alpha\alpha}_s(k_p,t)$ from $S^{\alpha\beta}(k)$.  Instead with our evolutionary strategy (sec.~\ref{sec:es}) we are able to construct the memory function $M^{\alpha\beta}(k_p,t)$ that produces the observed dynamics.}
\label{fig:sketch}
\end{figure*}

Understanding the dynamics of supercooled liquids approaching the glass transition represents one of the major challenges in condensed matter science~\cite{Anderson1995,Lubchenko2007,Berthier2011,Langer2014}. 
The most striking signature of this phenomenon is the dramatic increase in viscosity or relaxation time upon a relatively mild change in the thermodynamic control parameters. 
Despite the magnitude of this effect, there are no substantial changes in the microscopic structure of the material, which severely hinders our understanding of the mechanisms underlying the glass transition.

In recent years, machine learning algorithms have been successfully employed to capture subtle changes in the local structure of glassforming materials to create accurate predictors of the dynamics. The first such example is a machine-learned parameter called softness~\cite{Cubuk2015,Schoenholz2016,xiaoguang19,cubuk20,tah22}, which, based on support vector machines and physical intuition, identifies key structural features that strongly correlate with local particle dynamics.
Neural networks can also identify local structures~\cite{Boattini2018,Boattini2019} and correlate them to local dynamics~\cite{Boattini2020,oyama22}.
Furthermore graph neural networks~\cite{Bapst2020} have shown that the graph structure of each particle's local environment contains significant information to predict its long-time dynamics. It was later demonstrated that more refined observables could be calculated~\cite{Boattini2021} and combined with simpler models to capture the connection between statics and dynamics in glassy systems~\cite{Alkemade2022,jung22}.
Similar results can be achieved even with information theory~\cite{jack14,Paret2020} and dimensionality reduction~\cite{Paret2020,coslovich22}.
Recently, neural networks have also been used to find complex order parameters for glassy dynamics~\cite{D2SM00310D}. 
However all these approaches are based on a particle resolved description of the system, which requires knowledge of the location of every single particle and its precise local environment. Unfortunately these particle resolved quantities are not always easy to measure, and furthermore, single-particle properties do not easily lend themselves to statistical-physical theory development. Hence, a more collective description is often preferred.

Here we propose an alternative approach that is not based on local single-particle features, but instead on system averaged quantities. 
This approach takes inspiration from collective theories of the glass transition~\cite{Gotze1992,Xia1999,Tarjus2005a,Sausset2008,Biroli2012,Dell2015,Rizzo2015,Rizzo2020,Janssen2015a,Ciarella2021b,liu2021,Szamel2022} that aim to predict the glassiness of the system using collective static observables that do not need to be resolved per particle. 
The cornerstone of our method consists of rewriting the dynamics of supercooled liquids following the Mori-Zwanzig procedure~\cite{Gotze1992,Reichman2005} to obtain a form of a generalized Langevin equation called the memory equation~\cite{Kob2002}.
From a mathematical point of view, this equation takes as input the statistically-averaged static structure of the system, mainly through the static structure factor $S(k)$ which is a function of the wave vector $k$. Using $S(k)$ as the initial boundary condition, the memory equation can then be used to predict the time-dependent dynamics of the system, quantified by the intermediate scattering function $F(k,t)$ at a given time $t$. 
The key bottleneck, however, is finding the exact memory function that governs the dynamics of $F(k,t)$; this memory function should account for the dynamical slowdown of supercooled liquids, but its functional form is a priori unknown. 
After decades of intense research, scientists have been able to solve only approximations of this equation, like mode-coupling theory (MCT)~\cite{Gotze1992,Franosch1997,Voigtmann2003,gotze2008complex,Reichman2005,SzamelPRL2003,brader09,Weysser2010,Janssen2018primer,Ciarella2021b,Debets2021}. 

In this paper, we use machine-learning to approximate the memory function. In particular we discuss two different approaches to the problem: (i) first we train a deep neural-network (DNN) in order to learn the memory equation using data measured from computer simulations of binary mixtures interacting via Lennard-Jones (LJ) or Weeks-Chandler-Andersen (WCA) potentials. In this approach the DNN plays the role of both the memory equation and the memory function itself.  (ii) Then we develop an evolutionary strategy tailored to identify and construct the memory function that produces the dynamics observed in the simulations.

In Fig.~\ref{fig:sketch} we sketch the concept of our approach.
First we collect extensive simulation data for LJ and WCA binary mixtures. For different values of temperature $T$ and density $\rho$ we measure the static structure factor $S^{\alpha\beta}$ and the self intermediate scattering function $F^{\alpha\alpha}_s$, where $\alpha$ and $\beta$ are the indexes to represent the species of the mixture. These averaged descriptors are then given to a DNN that we train to predict the intermediate scattering function for a given $S^{\alpha\beta}$, under the assumption introduced in section sec.~\ref{sec:dnn}. We show that our DNN achieves excellent performance, thus concluding that the network can learn the memory equation. Next, in order to obtain some physical intuition about the memory function we introduce an evolutionary strategy that, given the intermediate scattering function and the structure of the memory equation, is able to describe the memory function in a parametrized functional form.

Overall, the results of our model are twofold: we introduce an effective approach that is able to rapidly predict the collective dynamics of the system from static measurements, once the model has been trained. Secondly, we propose a representation 
of the memory function that may be more convenient, and perhaps more realistic, than some state-of-the-art theories. The function parametrized by our machine-learning algorithm can be informative for future efforts aimed at developing a more quantitative theory of the glass transition.
In the next sections we will discuss the mechanism of our machine-learning approach, how to generalize it to other systems, and some implications 
of our findings.

\section{Results}
\subsection{Dynamics of supercooled liquids from neural networks}
\label{sec:dnn}
We train a multi-layer perceptron (MLP) to predict the dynamics of supercooled liquids from static averaged quantities.
The MLP is a fully connected DNN with a simple feed-forward architecture.
The dynamics is characterized by measuring the intermediate scattering function $F^{\alpha\beta} (k,t)$ from the simulations (details in sec.~\ref{sec:sim}).
Rather than a full $k$-dependent description of the dynamics, we follow the standard procedure~\cite{Berthier2010,Janssen2018primer,Flenner2005} for binary mixtures, which consists in the following steps: (i) we focus only on $k=k_p\equiv |\bm{k}_\mathrm{peak}^{AA}|$ which is the location of the main peak of $S^{AA}$ and it is arguably the  most descriptive wavenumber for such supercooled mixtures~\cite{Kob2002}, (ii) we consider only the self part of the intermediate scattering function assuming that it also represents the collective dynamics, and (iii) we ignore the mixed term $F^{AB}(k,t)$ since $F^{\alpha\beta}_s=0$. Hence we end up describing the dynamics using $F_s^{\alpha\alpha}(k_p,t)$, where $\alpha=A,B$ represents the two species. 
 
Furthermore, we define $F_s^{\alpha\alpha}(k_p,t)$ in the following way:   
\begin{align}
\label{eq:normaliz}
    F_s^{\alpha\alpha}(k_p,t) =
    \frac{1}{N}\langle & \sum_i^N  e^{-i\bm{k}_p\bm{r}_i^\alpha(0)} e^{i\bm{k}_p\bm{r}_i^\alpha(t)}\rangle 
    \nonumber \\ & \cdot
    \underbrace{\frac{1}{N}\langle \sum_i^N e^{-i\bm{k}_p\bm{r}_i^\alpha(0)} \sum_j^N e^{i\bm{k}_p\bm{r}_j^\alpha(0)}\rangle }_{\equiv S^{\alpha\alpha}(k_p)},
\end{align}
where $N$ is the total number of particles and $\bm{r}_{i}^{\alpha}(t)$ is the position of particle $i$ of species $\alpha$ at time $t$.
Our choice of normalization imposes $F_s^{\alpha\alpha}(k_p,t=0)=S^{\alpha\alpha}(k_p)$. In our work, this has two advantages: we can now directly compare our prediction of $F_s^{\alpha\beta}(k_p,t)$ with collective theories like MCT under assumptions (i)-(iii), and secondly, our machine learning approach gives more importance to the most glassy (slowest) dynamics, because they are rescaled by a larger factor corresponding to their larger values of $S^{\alpha\beta}(k_p)$.

The MLP that we use to predict the dynamics takes as input the static structure factor $S^{\alpha\beta}(k)$ on a uniform grid of $N_k=100$ wave numbers in the range $0\le k\le 40\sigma_{AA}$, including the $\alpha\neq\beta$ terms, for a total of $300$ values of $S^{\alpha\beta}(k_i)$ (considering the $\alpha\beta=\beta\alpha$ symmetry).
In addition the MLP receives the temperature $T$, the density $\rho$, a label for the interaction type (i.e., either LJ or WCA), and the logarithm of the time at which it has to predict $F^{\alpha\beta}_s(k_p,t)$. 
With this input-output architecture after training we can easily tune the value of $t$ in the input to reconstruct the full time dependence. 

\begin{figure}
\centering
\includegraphics[width=\columnwidth]{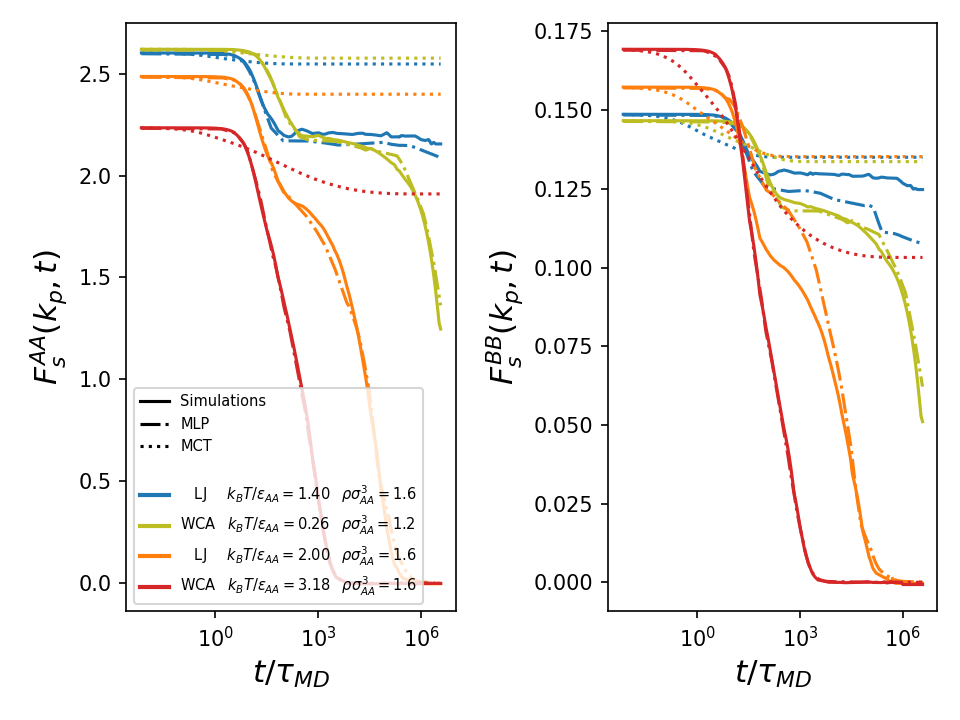}
\caption{Predictions of the multi-layer perceptron (MLP) for the self intermediate scattering function of binary mixtures, for the $AA$ and $BB$ components, normalized as defined in Eq.~\ref{eq:normaliz}. We report four state points from the test set (not used for training): the red is a warm liquid, the orange is a supercooled liquid, the yellow is a strongly supercooled liquid and blue is a glass.
We compare the MLP (dot-dashed lines) with simulations (solid lines) and mode-coupling theory (dotted line) predictions.
}
\label{fig:multi-f}
\end{figure}
In Fig.~\ref{fig:multi-f} we show that the MLP (detailed in sec.~\ref{sec:MLP}) produces good predictions in the full temperature range, from high-temperature liquid to glass on which it was trained.
In this figure we also report the results of mode-coupling theory (dotted lines), which is a theory based on the same static information used as input for the MLP.
The results indicate that the MLP significantly outperforms MCT in predicting the realistic dynamics.

\begin{figure}
\centering
\includegraphics[width=\columnwidth]{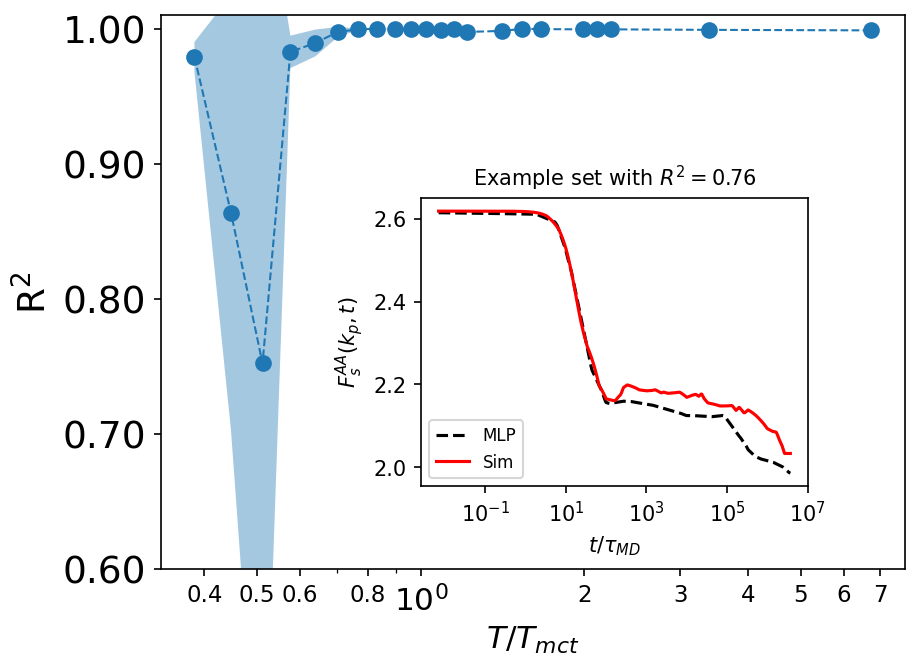}
\caption{Performance of the multi-layer perceptron (MLP) in predicting the self-intermediate scattering function of binary mixtures. We report the R$^2$ score as a function of temperature, normalized by $T_{mct}$ which is the temperature at which mode-coupling theory predicts the glass transition. 
Its value is averaged over states with similar values of $T/T_{mct}$ and the colored region represents the standard deviation of each bin. A value of $1$ represents perfect predictions. In the inset we compare the MLP prediction with the target simulation, for a set with R$^2=0.76$. We conclude that on average the model predictions are good across the entire temperature range. 
}
\label{fig:r2}
\end{figure}
To quantify the performance of the MLP we measure its predictivity on data outside of it training set. The model has been trained using $90\%$ of the data available while we use the remaining data to evaluate the R$^2$ score:
\begin{equation}
\mathrm{R}^2= 1 - \frac{SS_\mathrm{res}}{SS_\mathrm{tot}},
\end{equation}
where $SS_\mathrm{res}$ is the sum of squared residuals
\begin{equation}
    SS_\mathrm{res} = \sum_{\mathrm{observation}_i} \left( \mathrm{prediction}_i - \mathrm{truth}_i\right)^2
    \label{eq:ssres}
\end{equation}
and $SS_\mathrm{tot}$ is the total sum of squares
\begin{equation}
    SS_\mathrm{tot} = \sum_{\mathrm{observation}_i} \left( \mathrm{prediction}_i - \overline{ \mathrm{prediction}}\right)^2,
    \label{eq:sstot}
\end{equation}
where $\bar{...}$ represents the mean.
Notice that one observation in Eqs.~\ref{eq:ssres}-\ref{eq:sstot} corresponds to one time $t=t_i$ of $F_s(k_p,t)$ for any given $T$ and $\rho$, so the sum runs over all the temperatures $T$, densities $\rho$ and times $t_i$. Moreover, the train/test split is performed separating $10\%$ of the $\{$LJ/WCA$;T;\rho\}$ states rather than $10\%$ of all the data points, in order to avoid training the MLP using data strongly correlated with the test set.

\begin{table}
\tiny
\renewcommand{\arraystretch}{2.5}
\newcolumntype{z}{>{\centering\arraybackslash\columncolor[HTML]{bfbfbf}}m{1cm}}
\newcolumntype{g}{>{\centering\arraybackslash\columncolor{white}}m{1.1cm}}
\resizebox{\columnwidth}{!}{%
\begin{tabular}{ |z|g|g|g|g| }
\hline
\rowcolor{lightgray} \cellcolor[HTML]{8acf71} $\frac{k_BT_{mct}}{\epsilon_{\scaleto{AA}{2pt}}}$ 
&$\rho\sigma_{\scaleto{AA}{2pt}}^{\scaleto{3}{2pt}}=1.2$
&$\rho\sigma_{\scaleto{AA}{2pt}}^{\scaleto{3}{2pt}}=1.4$
&$\rho\sigma_{\scaleto{AA}{2pt}}^{\scaleto{3}{2pt}}=1.6$
&$\rho\sigma_{\scaleto{AA}{2pt}}^{\scaleto{3}{2pt}}=1.8$ \\
\hline
 WCA &
 0.74 &
 1.77 &
 3.49 &
 5.10 \\
\hline
 LJ &
 0.90 &
 1.87 &
 3.53 &
 5.10\\
\hline
\end{tabular}
}
\caption{\label{tab:Tmct} Temperatures that we use to normalize the data. The values of $k_BT_{mct}/\epsilon_{AA}$ correspond to the critical temperature of mode-coupling theory for the WCA and LJ mixtures~\cite{Voigtmann2008,Berthier2010,Berthier2011}, below which fit-parameter-free MCT predicts that the system is a glass. }
\end{table}

A value of R$^2=1$ corresponds to a perfect fit, so R$^2$ is an effective measure to evaluate the quality of the model.
In Fig.~\ref{fig:r2} we report the R$^2$ score as a function of $T/T_{mct}$, where $T_{mct}$ is the temperature at which fit-parameter-free mode-coupling theory predicts the glass transition, reported in Tab.~\ref{tab:Tmct}. With this normalization we can average simulations at different densities and different pairwise interactions to produce a single curve.
In this work we are not interested in the precise estimation of the critical point of MCT for our mixtures, but a more detailed discussion is available in Ref.~\cite{Voigtmann2008,Berthier2010,Berthier2011}. 
In the supplementary information we show the root mean squared error (RMSE) for the same data. 
The trend of the curve shows that the performance of the MLP starts to drop at $T<T_{mct}$, when the two-step relaxation becomes more prominent and thus the intermediate scattering function is a more 'complex' function with more features to predict.
At $T<0.6 T_{mct}$ the MLP predictions exhibit pronounced fluctuations, but the predictions are still good on average, as we highlight in the inset of Fig.~\ref{fig:r2}. 
Surprisingly at even lower temperatures, when the dynamics is even slower, the performances increase, approaching again the perfect R$^2=1$ score. This is a consequence of the fact that in the glass the second relaxation happens at $t>t_\mathrm{max}$, which is outside the time window that the model observes, so $F$ is characterized by a single relaxation, thus being easier to learn for the model.

\subsubsection{Transferability}    
\begin{figure}
\centering
\includegraphics[width=\columnwidth]{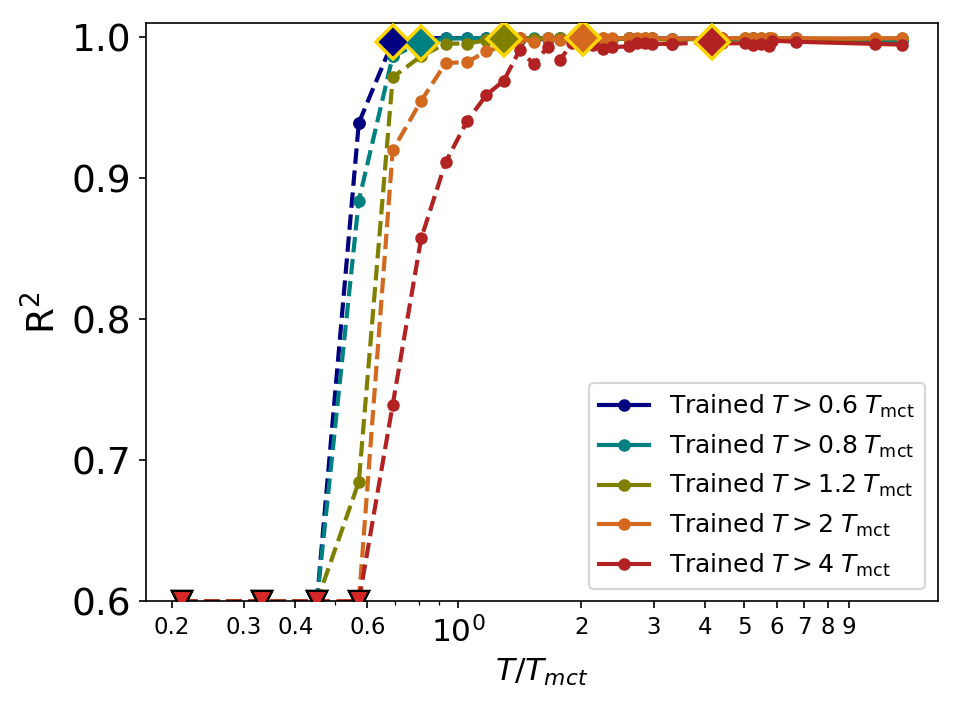}
\caption{Temperature transferability of the MLP: we report the R$^2$ score as a function of the temperature. For each curve the model is trained using only temperatures larger than the one corresponding to the rhombus. The solid lines represent the regions in which the model has been trained, while it has never seen data from the dashed line region.
}
\label{fig:Ttransf}
\end{figure}
\textbf{Temperature transferability:} One of the main strengths of machine learning is the possibility to train a model in a more favorable situation (e.g.\ when there is more data available) and deploy it in a less favorable one.
For computer simulations the hardest region to sample is the low temperature regime, because the dynamics is much slower and more time is required to sample all relevant structural rearrangements. 
This means that it is much easier to collect data at high $T$ than at low $T$.

Here we show the temperature transferability of our model, i.e.\ its performance at temperatures different from the training.
We report in Fig.~\ref{fig:Ttransf} the R$^2$ score when the MLP is trained only using data for $T>T_0$.
The RMSE is reported in the supplementary information.
As expected the model performs excellently in its training region (solid lines), but outside (dashed lines) its score starts dropping.
The results of Fig.~\ref{fig:Ttransf} suggest the existence of two regimes:
(i) for $T>0.8\,T_{mct}$ we have full transferability. It is in fact possible to train the model at high temperature and retain good predictions, as evidenced by the red curve in Fig.~\ref{fig:Ttransf}.
(ii) When $T<0.8\,T_\mathrm{mct}$ the transferability is restricted. In Fig.~\ref{fig:Ttransf} we show that the best we can do is to transfer the training at $T>1.2\,T_\mathrm{mct}$ down to $T> 0.5 T_\mathrm{mct}$.

Overall we can conclude that our approach is transferable in the (i) MCT regime corresponding to $T>0.8\,T_\mathrm{mct}$, while the transferability is very limited in the (ii) $T<0.8\,T_\mathrm{mct}$ activated regime. This low temperature region is in fact characterized by the appearance of heterogeneous activated dynamics and facilitation, that become dominant at low temperature~\cite{guislein22,scalliet22}.
We hypothesize that this different relaxation mechanism is the reason behind the lack of transferability.
In summary, except for the very low temperature region, our approach presents some degree of temperature transferability that we can use to reduce the need for data that are harder to collect.


\begin{figure}
\centering
\includegraphics[width=\columnwidth]{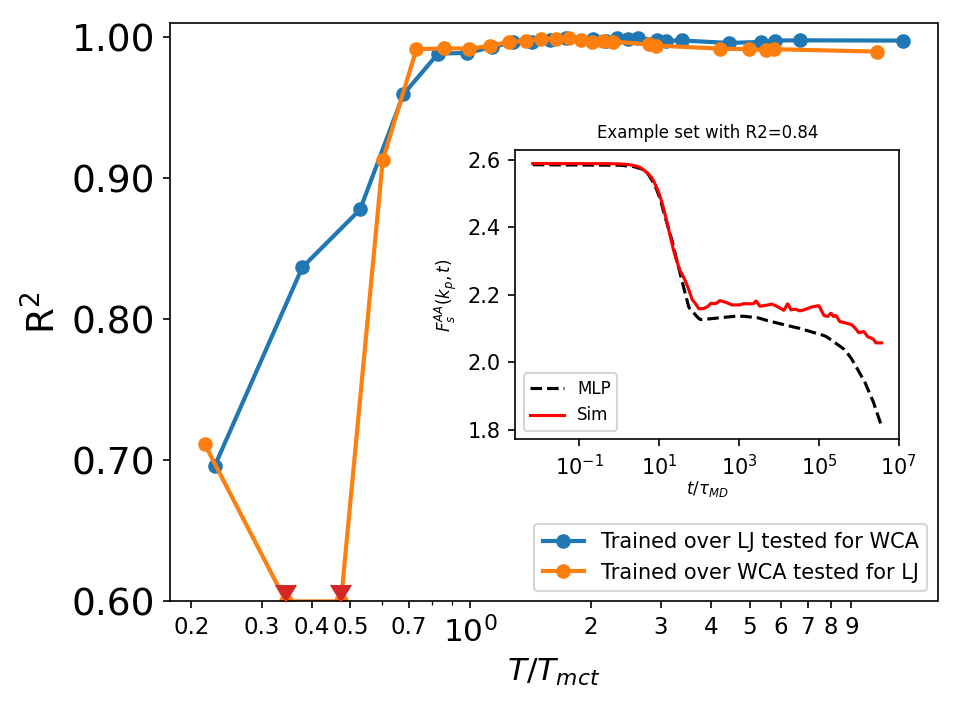}
\caption{Model transferability of the MLP. We report the R$^2$ score of the model when tested for a different interaction potential not included in the training. 
The inset qualitatively shows that R$^2>0.8$ corresponds to acceptable predictions. 
}
\label{fig:modeltransf}
\end{figure}
\textbf{Model transferability:} We show in this section the performance of the MLP in predicting the dynamics of computer simulations produced with a different interaction than the one the MLP has been trained on.
Our library consists of simulation data for binary LJ and binary WCA, so their dynamics is similar at high temperature and/or high density, but it becomes significantly different approaching $T_\mathrm{mct}$.
In Fig.~\ref{fig:modeltransf} we show the prediction of the MLP when it is trained over LJ and tested over WCA and vice versa, reporting the R$^2$ score as a function of normalized temperature.
We see that at $T>0.6 T_{mct}$ the MLP produces excellent predictions even when it is trained for a different interaction potential.
However the quality drops for very low temperature, when minute differences between the LJ and WCA  structures are amplified to enormous differences in dynamics~\cite{Berthier2010,Berthier2011a,Coslovich2012,Landes2020,Ciarella2021b,Debets2021} and configurational entropy~\cite{Banerjee2014,nandi15,banerjee20}.
In particular the model performs poorly when it is trained using the WCA potential and tested over the LJ data.
This is a consequence of the fact that the LJ model becomes a glass at a higher temperature compared to the WCA model~\cite{Berthier2010,Berthier2011a,Ciarella2021b}, so there is a region where the LJ model is infinitely slower than the WCA.

Overall we see that it is possible to use data measured from another system to obtain reliable predictions for a different (but relatively similar) system. However those predictions become unreliable at very low $T$, when approaching the glass transition. 

\subsection{Evolutionary strategy for the memory function}
\label{sec:es}
\begin{figure}
    \centering
    \includegraphics[width=\columnwidth]{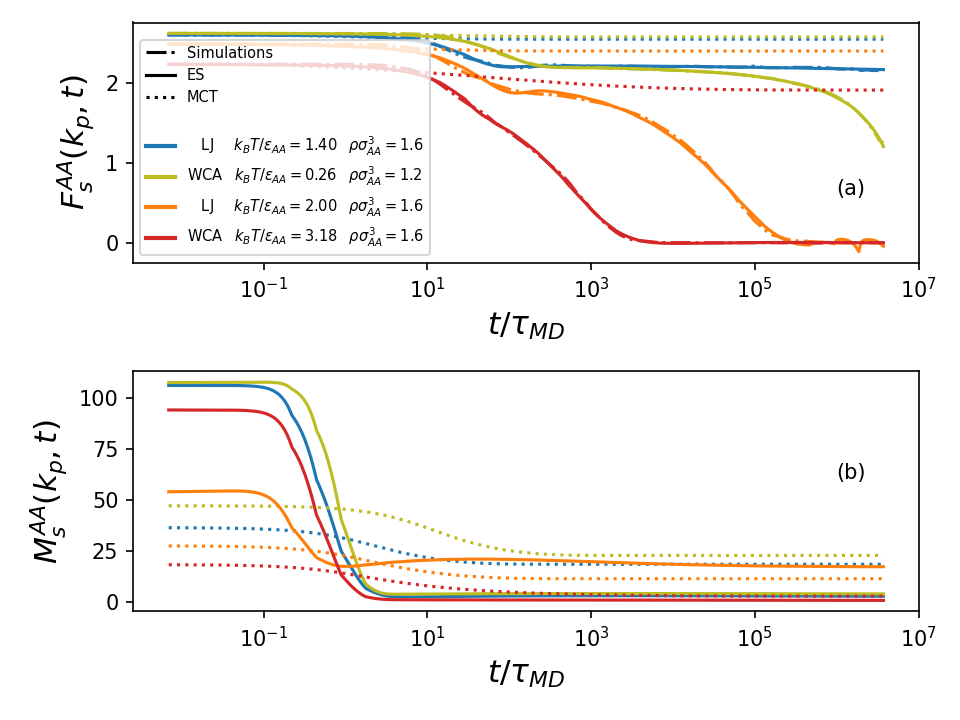}
    \caption{Predictions of the evolutionary strategy (ES) for the memory function of binary mixtures, with focus on the dominant $AA$ component at $k=k_p$. We report four state points at different temperature: the red is a warm liquid, the orange is a supercooled liquid, the yellow is a strongly supercooled liquid and blue is a glass.
    In (a) we see that the ES (solid) perfectly reproduces the simulations (dot-dashed). In panel (b) we report the memory that the ES uses to reproduce the simulated $F_s^{AA}$ (solid) and we compare it to MCT (dotted) which produces instead very bad predictions.
}
    \label{fig:FM-es}
\end{figure}

In the previous section we have seen how to numerically correlate static structural information with the dynamics of the system.
Such a deep learning approach however does not tell us much about the physics of the system.
Here we employ a physics-informed strategy that is created in order to avoid the application of a \textit{black box}~\cite{Rudin2019,Murdoch2019}, but instead we bound the machine learning model to play the role of a single physical unknown function: the memory function.

In order to develop this physics-informed strategy we start from the exact memory equation that describes the overdamped dynamics of liquids~\cite{Kob2002}:
\begin{align}
\frac{d F^{\alpha\beta}_s(k_p, t)}{d t}&+\Omega^2_{\alpha\beta}(k_p) F^{\alpha\beta}_s(k_p, t)+\nonumber \\ & \int_{0}^{t} d \tau M^{\alpha\gamma}(k_p, t-\tau) \frac{\partial F^{\gamma\beta}_s(k_p, \tau)}{\partial \tau}=0,
\label{eq:gle}
\end{align}
where $\Omega^2_{\alpha\beta}(k_p)$ is a constant representing the vibrational term~\cite{Gotze1992,Reichman2005}.
While the equation above is formally exact, unfortunately the memory function $M^{\alpha\beta}(k_p, t-\tau)$ that appears in the integral is unknown, and thus the equation cannot be solved. MCT is based on an uncontrolled approximation of this memory function that lead to excellent semi-quantitative results~\cite{Reichman2005,sciortino2001debye,Weysser2010}, but its predictions are not able to exactly capture the full phenomenology of the glass transition~\cite{Berthier2010}.
While there are some ways to invert Eq.~\ref{eq:gle}~\cite{Baity-Jesi2019,Landes2020}, and various approaches to numerically estimate the memory function~\cite{Carof2014,Jung2017,Klippenstein2021,Vroylandt2022}, there is no consensus on the general form of $M$. More importantly, most of these procedures to calculate $M$ are very sensitive to noise, thus requiring refined data, and overall they are often computationally expensive.

Here we show that it is possible to effectively parameterize $M$ as a sum of stretched exponentials, whose coefficients can be determined by an evolutionary algorithm.
In choosing this exponential representation, we have drawn inspiration from theories such as MCT~\cite{Gotze1992,Bengtzelius1984} for which the known (albeit approximate) memory function typically has a similar structure as the intermediate scattering function, combined with the fact that $F_s(k,t)$ is known experimentally to behave as a stretched exponential for long times~\cite{vanmegen93}. 
Thus, we represent the memory function in the following way:
\small
\begin{align}
    M^{\alpha\beta}(k,t) &= c^{\alpha\beta}(k) + \sum_{i=0}^{N_{exp}} w^{\alpha\beta}_i(k) \;\times \nonumber \\& \left\{ \left[1-a^{\alpha\beta}_i(k)\right] e^{-\left(\frac{t}{\tau^{\alpha\beta}_i(k)}\right)^{b^{\alpha\beta}_i(k)} } +a^{\alpha\beta}_i(k)\right\}
    \label{eq:strexp}
\end{align}
\normalsize
which is a sum of $N_{exp}$ stretched exponentials, for a total of $3(1+4N_{exp})$ parameters for every state.
Notice that our parametrization also allows for standard or compressed exponentials, according to the value of $b_i^{\alpha\beta}(k)$.
As previously discussed (sec.~\ref{sec:dnn}), we focus only on $k=k_\mathrm{peak}^{AA}$, which is usually the most important wavevector in simple glassformers such as the systems studied here. 

The details of our evolutionary strategy (ES) are discussed in sec.~\ref{sec:method-es}. Briefly, the ES is an optimization technique inspired by evolution and natural selection. It works by creating new generations of memory functions (parameterized as Eq.~\ref{eq:strexp}) by cross-breeding the previous generation and adding random mutations. For each of them we then solve the memory equation (Eq.~\ref{eq:gle}) obtaining $\left\{F^{\alpha\alpha}_{ES}(t_i) \right\}$ and then we do one step of evolution favoring the individuals that minimize the following loss function:
\begin{equation}
    \mathcal{L}_{ES}=\sum_{t_i,\alpha} \left[ F^{\alpha\alpha}_s\left(k_P,t_i\right) - F^{\alpha\alpha}_{ES}(t_i)\right]^2 +\mathrm{regularization}
    \label{eq:loss}
\end{equation}
which is the squared difference between predicted and observed dynamics. In the Supplementary Information we verify that the ES solutions are physically reasonable by replacing the intermediate scattering function with the MCT approximation $F^{\alpha\alpha}_s\left(k_P,t_i\right)\xrightarrow{}F^{\alpha\alpha}_{MCT}(k_p,t_i)$ and confirming that the evolutionary strategy converges towards $M^{\alpha\beta}_{MCT}(k_p,t)$. Notice however that Eq.~\ref{eq:strexp} allows unrealistic short time non-monotonic behavior, that we observe sometimes. Overall, we believe that our ES solutions still retain a fair degree of realism.

The results of the ES are reported in Fig.~\ref{fig:FM-es}.
In panel (a) we show that the ES is able to recover the simulated self intermediate scattering function  $F^{\alpha\alpha}_s\left(k_P,t_i\right)$ that we are targeting. The figure presents four curves that range from the liquid to the glass phase. We have also verified that the ES solution converges to the simulated $F^{\alpha\alpha}_s(k_p,t)$ for any temperature and density that we have available. Even though we have only focused on $k=k_p$, we speculate that our approach should work for any value of $k$ since the structure of Eqs.~\ref{eq:gle}-\ref{eq:strexp} does not depend on $k$. Note however that, unlike MCT, this parametrized memory function does not contain explicit couplings among different wave numbers, nor any self-consistent feedback mechanism to drive dynamical slowdown.

In Fig.~\ref{fig:FM-es}(b) we report the results for the memory function.
Following the parametrization introduced in Eq.~\ref{eq:strexp} the ES is able to converge to memory functions that produce the realistic $F^{\alpha\alpha}_S(k_p,t)$ of Fig.~\ref{fig:FM-es}(a).
In particular the memory functions presented in Fig.~\ref{fig:FM-es}(b) are obtained setting $N_{exp}=2$. 
We highlight the differences between the $M$ at which the ES converged, with the MCT approximation (dashed lines), that unsurprisingly overestimates the glassiness of these systems~\cite{Kob2002,Berthier2010,Janssen2018primer}.
We conclude that the ES method we introduced is an effective way to obtain a realistic memory function.

\begin{figure}
\centering
\includegraphics[width=\columnwidth]{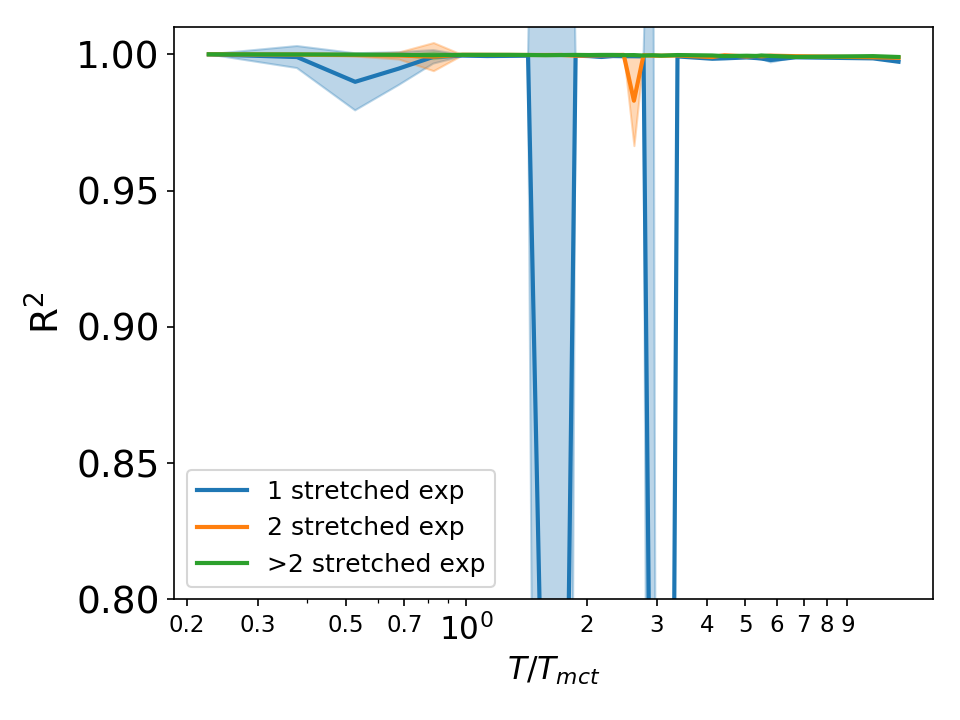}
\caption{Performance of the evolutionary strategy in finding the memory function that correctly predicts the dynamics. We parameterize the memory following Eq.~\ref{eq:strexp}, while the different curves correspond to different values of $N_{exp}$. We show that $N_{exp}=2$ is the minimum number of exponentials to achieve good performances, which corresponds to a total of $27$ parameters.}
\label{fig:es-r2}
\end{figure}

Lastly we discuss the physical insights that we can harness from our evolutionary approach. 
We have seen that the ES converges to the simulated intermediate scattering function if we model the memory function following Eq.~\ref{eq:strexp}.
Since Eq.~\ref{eq:gle} is exact, we know that the real memory function has to produce the same output as our ES when propagated through Eq.~\ref{eq:gle}. This means that we can look at the structure of our ES solutions to have some intuition about the real memory.
In particular, in Fig.~\ref{fig:es-r2} we report the R$^2$ for different parametrizations containing a different number of stretched exponentials and we verify that a good solution needs at least $N_{exp}=2$. 
In the supplementary information we also show that the second stretched exponential becomes relevant only approaching the glass transition, in the same range where localized unstable modes become relevant~\cite{Coslovich2019}, and when energy-driven and entropy-driven activation start to compete~\cite{Carbone2022}.
We also know that activation and facilitation become relevant only at very low temperature~\cite{guislein22,scalliet22}.
This suggests that there are at least two separate relaxation channels that real materials follow while relaxing, with the second channel becoming dominant below $T_{mct}$, i.e.\ the temperature that is often identified with a crossover~\cite{Andersen05}. It is important to recall however that the $T_{mct}$ used in our work is obtained from fit-parameter-free MCT, while other works have considered different definitions of $T_{mct}$, and hence these comparisons should be treated with caution.

Moreover, the results derived from the evolutionary strategy may also provide cues for analytical modeling. 
Models like MCT invoke uncontrolled approximations to solve Eq.~\ref{eq:gle}. One possibility is the exponential closure that assumes that $M^{\alpha\beta}(k,t)\sim \exp\left[-\Omega^2_{\alpha\beta}(k) t\right]$.
This schematic model is simple enough to be solved analytically, but its results are not satisfying~\cite{Leutheusser1984,Bengtzelius1984,Janssen2015a,Ciarella2019pnas,CiarellaPhdThesis}.
Inspired by our ES we may propose a double stretched-exponential memory defined as 
\begin{equation}
    M^{\alpha\beta}(k,t)= \exp\left[-\Omega_{\alpha\beta}^2(k) t^{\nu} \right] +  K^{\alpha\beta}(k)\exp\left[-\tilde{\Omega}_{\alpha\beta}^2(k) t^{\mu} \right].
    \label{eq:mctclosure}
\end{equation}
This schematic model retains a structure of Eq.~\ref{eq:gle} similar to the exponential closure, but according to our ES results it should be  more appropriate to describe the glassy dynamics, provided that it is properly parametrized. Note however that the correct parametrization would still require numerical fitting (e.g.\ via an ES). Nonetheless, we argue that our ES strategy can be used to explore different functional forms that may improve upon the conventional MCT closure approximation. 

In summary we have defined here a simple approach to determine the memory function $M$ given the self-intermediate scattering function $F_s$. The ES is fast and reliable across all the temperatures and densities studied. The results suggest that at least two relaxation channels need to be considered, giving rise to the machine-learning-inspired schematic model of Eq.~\ref{eq:mctclosure}.

\section{Discussion}

The goal of this study was to show how a neural network can understand and predict the dynamics of supercooled liquids from static information.
Our approach is based on averaged quantities such as static structure factors that are easier to measure experimentally, compared to particle resolved ones. We are also able to interpret the mechanism of the machine learning model to gain some physical intuition about the glass transition.

In Fig.~\ref{fig:multi-f}-\ref{fig:r2} we show that we can train a multi-layer perceptron to efficiently predict the dynamics (represented by the self intermediate scattering function $F_s^{\alpha\beta}$), using only statistically-averaged static information.
This implies that dynamical information is encoded in the static structure of the system, similarly to how higher-order correlations are partially encoded into two body structure~\cite{Coslovich2012}.
Furthermore since the main input of our neural network is the static structure factor $S^{\alpha\beta}$, our results corroborate the idea that two body static correlations when elaborated using an expressive approach like our MLP, are enough to describe the dynamics. 

One of the main problems in studying systems close to the glass transition is data collection, because experiments and simulations at deeply supercooled temperatures are slow. We show that our MLP performs relatively well when only high temperature data are provided for training (Fig.~\ref{fig:Ttransf}) or when the MLP is used to make predictions for different variations of the model that is used for training (Fig.~\ref{fig:modeltransf}). Unfortunately this transferability drops when the system is approaching the glass transition, because we know that minute changes in the structure correspond to enormous changes in the dynamics. 
Overall this means that, if the interest is in the liquid regime it is possible to fully exploit the MLP transferability by training the model where data is easily available, but to accurately describe the glassy regime, glassy data is actually required.

We then develop a physics-inspired method to obtain an expression for the memory function that realistically describe our data.
Instead of using deep learning as a black box to connect statics to dynamics, we rewrite the dynamics as a memory function and we replace this memory with our machine learning model.
Our approach circumvents inverse Laplace transforms, which can be computationally expensive, by using an evolutionary strategy that parametrizes a pre-defined functional form for the memory function.
We show in Fig.~\ref{fig:FM-es} that the ES easily converges to memory functions that reproduce the real dynamics observed in simulations.

Our physics-inspired evolutionary approach is also able to give some intuition about the physics. The results in Fig.~\ref{fig:es-r2} let us conclude that the memory function can be effectively parameterized as a sum of two stretched exponentials.
We can interpret those two stretched exponentials as two relaxation processes that describe the complex multiscale relaxation of the glassy liquid, and we can also see that only one is needed to describe the liquid phase (SI Fig.~\ref{fig:2strexp}). One possible interpretation for the existence of two dominant relaxation channels might be dynamic heterogeneity, i.e.\ the coexistence of transiently fast and slow groups of particles, which is known to emerge at temperatures below $T_{mct}$~\cite{Schoenholz2016,Boattini2020,guislein22}. However, more work is needed to identify the microscopic physical origins of the two stretched exponentials and to pinpoint the temperature regime where this effect is relevant.
This intuition motivates us to explore a schematic model where the memory function is exactly represented as two stretched exponentials, but we leave that for future work. 

In conclusion, we have introduced two data-driven tools to evaluate and describe the dynamics of supercooled liquids. The neural network that we propose can be efficiently trained and deployed to predict the self-intermediate scattering function from averaged quantities which are simple to measure experimentally. Lastly we have discussed a way to obtain an effective memory function using an evolutionary strategy, concluding that the memory can be reasonably represented as a sum of two stretched exponentials. We believe that our machine-learning method, once trained, can be efficiently applied to predict the dynamics of many other glass forming mixtures and that data-driven approaches to find suitable functional forms of the memory function may help guide the development of more effective theories to describe the glass transition.


\section{Methods}

\subsection{Computer simulations}
\label{sec:sim}
The models reported in this paper have been trained and tested using simulation data of two binary mixtures: the Kob-Andersen binary Lennard-Jones (LJ) mixture~\cite{Kob1994} and its Weeks-Chandler-Andersen truncation (WCA) ~\cite{wca1971}.
They are three-dimensional $80:20$ mixtures of particles $A:B$ interacting with each other via 
\begin{equation}
    V_{\alpha\beta}(r)=
    \begin{cases}
    4\epsilon_{\alpha\beta}\left[\left(\frac{\sigma_{\alpha\beta}}{r}\right)^{12}-\left(\frac{\sigma_{\alpha\beta}}{r}\right)^{6} +C_{\alpha\beta}\right],\hfill r\le r_{\alpha\beta}^c
    \\
    \mathrlap{\qquad 0}\hphantom{4\epsilon_{\alpha\beta}\left[\left(\frac{\sigma_{\alpha\beta}}{r}\right)^{12}-\left(\frac{\sigma_{\alpha\beta}}{r}\right)^{6} +C_{\alpha\beta}\right]},\hfill r>r_{\alpha\beta}^c
    \end{cases}
\end{equation}
where the cutoff radius $r^c_{\alpha\beta}$ is $2.5\sigma_{\alpha\beta}$ for LJ, and $r^c_{\alpha\beta}$ is $2^{1/6}\sigma_{\alpha\beta}$ for WCA~\cite{wca1971}. The choice of $C_{\alpha\beta}$ secures that $V_{\alpha\beta}(r^c_{\alpha\beta})=0$.
As usual~\cite{Kob1994}, we set $\epsilon_{AB}/\epsilon_{AA}=1.5, \epsilon_{BB}/\epsilon_{AA}=0.5, \sigma_{AB}/\sigma_{AA}=0.8, \sigma_{BB}/\sigma_{AA}=0.88$.

For these mixtures we perform molecular dynamics simulations in the $NVE$ ensemble using HOOMD-blue~\cite{anderson2008general}. First we equilibrate the system at different densities $\rho\sigma_{AA}^3 \in \lbrack 1.2,1.8\rbrack$ and temperatures $k_BT/\epsilon_{AA} \in \lbrack 0.2, 15 \rbrack.$  We impose periodic boundary conditions and set the box at $L=10\sigma_{AA}$ while tuning the density by changing the number of particles $N\in \left[1200,2000\right]$. All the particles have the same mass $m$. We run the simulations for $10^8$ timesteps of size $dt=10^{-3}\tau_{MD}$ with $\tau_{MD}=\left( m\sigma_{AA}^{2}/48\epsilon_{AA}\right)^{1/2}$, which allows us to sample up to $t_\mathrm{max}=3\cdot 10^6\tau_{MD}$. Notice that the trajectories that are deeply in the glass regime (i.e. where $F_s^{\alpha\alpha}(k,t\xrightarrow{}\infty) >0$) cannot be fully equilibrated due to the ergodicity breaking that defines the glass transition. 
Additionally, even though the mixture is designed to avoid crystallization, it is still possible for some specific trajectories to crystallize; in that case we remove such occurrences from our data.
Finally, we use the simulated trajectories to calculate the partial static structure factors $S^{\alpha\beta}(k)$ and the self intermediate scattering functions $F_s^{\alpha\beta}(k,t)$.

\subsection{MLP}
\label{sec:MLP}
We train a multi-layer perceptron to predict the dynamics of supercooled mixtures from static information. 
The results reported in sec.~\ref{sec:dnn} are produced by a multi-layer perceptron. 
The network consists of 5 hidden layers of size $\left\{500, 400, 400, 300, 200\right\}$, interposed by ReLU activation functions, that transform the $304$ input features $\left\{S^{AA}(k_1),...,S^{BB}(k_{100}),T,\rho,\mathrm{interaction},t\right\}$ into the output $\left\{F_s^{AA}(k_p,t) ,F_s^{BB}(k_p,t)\right\}$. The MLP is trained for $1000$ iterations, taking approximately $1$ hour on a standard Intel i7 CPU. We also use $L1$ loss and Lasso regularization~\cite{bengio12} with the adam optimization algorithm~\cite{kingma16}.

\subsection{Memory equation and MCT}
In this paper we numerically solve Eq.~\ref{eq:gle}, which is called the memory equation~\cite{Kob2002}. This is an equation of the same class as the generalized Langevin equation (GLE) and we believe that out method can be tailored to solve a wide category of GLE equations with a structure similar to Eq.~\ref{eq:gle}, using physical intuition.

In general, it is possible to exactly describe the dynamics of liquids by deriving an expression for  
\begin{align}
    F^{\alpha\beta}(k,t) =
    \frac{1}{N}\langle & \sum_i^N  e^{-i\bm{kr}_i^\alpha(0)} \sum_j^N e^{i\bm{kr}_j^\beta(t)}\rangle,
\end{align}
from the solution of the overdamped equation
\begin{align}
\label{eq:mct}
\frac{d F^{\alpha\beta}(k, t)}{d t}&+\Omega^2_{\alpha\beta}(k) F^{\alpha\beta}(k, t)+\nonumber \\ & \int_{0}^{t} d \tau M^{\alpha\beta}(k, t-\tau) \frac{\partial F^{\alpha\beta}_s(k, \tau)}{\partial \tau}=0,
\end{align}
but unfortunately the memory function $M^{\alpha\beta}(k, t)$ is unknown.
Notice that in this paper we are mainly interested in solving Eq.~\ref{eq:gle} which corresponds to Eq.~\ref{eq:mct} if $F^{\alpha\beta}(k,t)$ is replaced with $F_s^{\alpha\alpha}(k_p,t)$.
We will also assume that the memory function is the same in Eq.~\ref{eq:gle} and Eq.~\ref{eq:mct}.


The memory equation can only be solved using some approximations like mode-coupling theory~\cite{Gotze1992,Voigtmann2003,Reichman2005,gotze2008complex,Ciarella2021b,Debets2021}, for this reason we also refer to the MCT equation.
The MCT approximation applied to Eq.~\ref{eq:gle} consists of the following definition of the memory function:
\begin{align}
\label{eq:Mmct}
M^{\alpha\beta}_\mathrm{mct}(k,t)=\frac{1}{2k^2}\frac{\rho}{x_\alpha x_\beta} & \sum_{\substack{\alpha'\beta'\\\alpha''\beta''}} \int \frac{d^3\bm{q}}{\left(2\pi\right)^2} \cdot \nonumber \\ & \cdot V_{\alpha\alpha'\alpha''}\left(\bm{q},\bm{k},\bm{k}-\bm{q}\right)F_s^{\alpha'\beta'}(k,t) \cdot \nonumber \\ & \cdot F_s^{\alpha''\beta''}(k,t) V_{\beta\beta'\beta''}\left(\bm{q},\bm{k},\bm{k}-\bm{q}\right),
\end{align}
where $x_\alpha=N_\alpha/N$ is the density of species $\alpha$ and the vertex function corresponds to
\begin{equation}
V_{\alpha\beta\gamma}\left(\bm{k},\bm{q},\bm{p}\right)=\left(\hat{k}\cdot\bm{q}\right)c^{\alpha\beta}(q)\delta_{\alpha\beta} + \left(\hat{k}\cdot\bm{p}\right)c^{\alpha\gamma}(p)\delta_{\alpha\gamma}, 
\end{equation}
with $S^{-1}_{\alpha\beta}(k)=\delta_{\alpha\beta}/x_\alpha - \rho c_{\alpha\beta}(k)$.

So overall, the inputs for the MCT equation are the bulk density $\rho$, the temperature $T$, and the structure factor $S^{\alpha\beta}(k)$.
In our numerical solution of the MCT equation we follow all the steps discussed in Ref.~\cite{Flenner2005,Weysser2010,Ciarella2021b,Debets2021} over a grid of $N_k=100$ points. 
As an alternative to the MCT approximation, we also solve Eq.~\ref{eq:gle} using Eq.~\ref{eq:strexp} instead of $M^{\alpha\beta}_\mathrm{mct}$, from the same inputs.
In summary, at any given $(T,\rho)$ we only require $S^{\alpha\beta}(k)$ as input to predict the microscopic relaxation dynamics of the system, either with MCT or with our parametrization of the memory identified by the evolutionary strategy.

\subsection{Evolutionary strategy}
\label{sec:method-es}
Evolutionary strategy (ES) is a class of machine learning optimization algorithms that are inspired by natural evolution in the following way: at every iteration (or generation), a population of parameters (the genotypes)  are perturbed by cross-breeding and mutations and their objective function (fitness) is evaluated.
Then the highest scoring parameters are recombined to populate the next generation, iteratively until the objective function is optimized.
The huge advantage of this class of algorithms is that they do not require back-propagation, which is particularly useful when the objective function is a complex integro-differential equation like our GLE (Eq.~\ref{eq:gle}).

We use covariance matrix adaptation evolution strategy~\cite{Hansen2001} (CMA-ES), a widely known method of the ES class which describes the population by a multivariate Gaussian.
The algorithm is available as a python package~\cite{hansen2019pycma}. Another advantage is that it does not require any hyper-parameter except for the initial condition and the population size represented by the initial variance of the gaussian $\sigma$. 
We then use CMA-ES to find the best function $M^{\alpha\beta}_{ES}$ parametrized as Eq.~\ref{eq:strexp} that reproduces the real dynamics. 
In practice at each step we propagate $M^{\alpha\beta}_{ES}$ through Eq.~\ref{eq:gle} and we compare its output $F^{\alpha\alpha}_{ES}(t)$ to the real dynamics $ F^{\alpha\alpha}_S\left(k_p,t\right)$, evaluating the fitness function $\mathcal{F}=-\mathcal{L}$ (Eq.~\ref{eq:loss}) for the evolution.
We use as initial condition $M=M_{mct}$ and for the initial population we set $\sigma=5\cdot 10^{-4}$. 
We use Lasso regularization to evaluate Eq.~\ref{eq:loss}.
Usually the procedure converges below an error $\epsilon<10^{-6}$ within a couple of hours of evolution.
Results in Fig.~\ref{fig:es-r2} show that the $F$ predicted by ES is very close to the real dynamics, so we conclude that $M$ is well represented by two stretched exponentials.

\section{Data availability}
The simulation results and the data analysis that support the findings of this study are available upon request from the corresponding author.

\section{Code availability}
The codes used to support the findings of this study are available upon request from the corresponding author.

\section{Acknowledgements}
We thank Ilian Pihlajamaa and Vincent Debets for their careful feedback and useful suggestions related to this work. This work has been financially supported by the
Dutch Research Council (NWO) through a START-UP grant
(LMCJ) and Vidi grant (LMCJ).

\section{Author contributions}
S.C. performed the simulations.
S.C. developed the MLP model.
S.C., M.C., E.B. developed the ES.
S.C. wrote the paper with the help of M.C., E.B., M.D. and L.M.C.J.
M.D., L.M.C.J. provided resources and supervision.    

\section{Additional information}
\textbf{Supplementary information} is available for this paper at link.

\clearpage

\newcommand{\beginsupplement}{
        \clearpage
        \setcounter{section}{0}
        \renewcommand{\thesection}{S\arabic{section}}
        \setcounter{equation}{0}
        \renewcommand{\theequation}{S\arabic{equation}}
        \setcounter{table}{0}
        \renewcommand{\thetable}{S\arabic{table}}
        \setcounter{figure}{0}
        \renewcommand{\thefigure}{S\arabic{figure}}
}
\beginsupplement

\section*{Supplementary information}
\subsection*{Root mean squared error}
\begin{figure}
\centering
\includegraphics[width=\columnwidth]{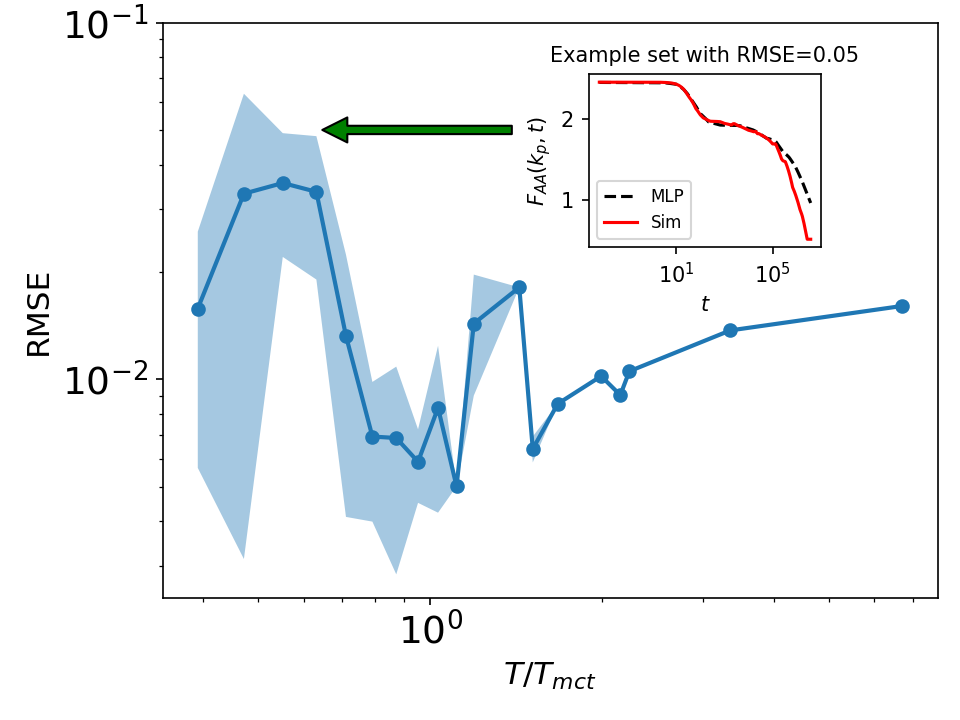}
\caption{Performance of the multi-layer perceptron (MLP) in predicting the self intermediate scattering function of binary mixtures. We report the RMSE as a function of temperature, normalized by $T_{mct}$. The RMSE is averaged over a discretized grid of states with similar values of $T/T_{mct}$ and the colored region represents the standard deviation of the bin. In the inset we compare the MLP prediction with the target simulation, for a set with RMSE$=0.05$ in order to qualitatively quantify the RMSE.
}
\label{fig:rmse}
\end{figure}
In Fig.~\ref{fig:rmse} we report the performance of the MLP (detailed in sec.~\ref{sec:MLP}). The model has been trained using $90\%$ of the data available and then we use the remaining test set to evaluate the root mean square error
\small
\begin{equation}
\mathrm{RMSE}=   \sqrt{\frac{\sum\limits_{i=1}^{N_\mathrm{test}}\sum\limits_{t_x=0}^{t_\mathrm{max}}\sum\limits_{\alpha=A,B} \left[ F^{\alpha\alpha}_\mathrm{sim}(k_p,t_x) -  F^{\alpha\alpha}_\mathrm{MLP}(k_p,t_x)\right]^2 } {N_\mathrm{test}}}.
\label{eq:rmse}
\end{equation}
\normalsize
We report it as a function of $T/T_{mct}$, where $T_{mct}$ is the temperature at which mode-coupling theory predicts the glass transition. With this normalization we can average simulations at different densities and different pairwise interactions to produce a single curve.
The trend of the curve shows that the MLP performs worse at $T<T_{mct}$, when the two-step relaxation becomes more prominent and thus the intermediate scattering function is a more 'complex' curve with more features to predict. 
Surprisingly the performance improves at very low temperature, but this is a consequence of the fact that in the glass the second relaxation happens at $t>t_\mathrm{max}$, which is outside of the observation window, so similarly to very high $T$, $F$ is characterized by a single relaxation.

\begin{figure}
\centering
\includegraphics[width=\columnwidth]{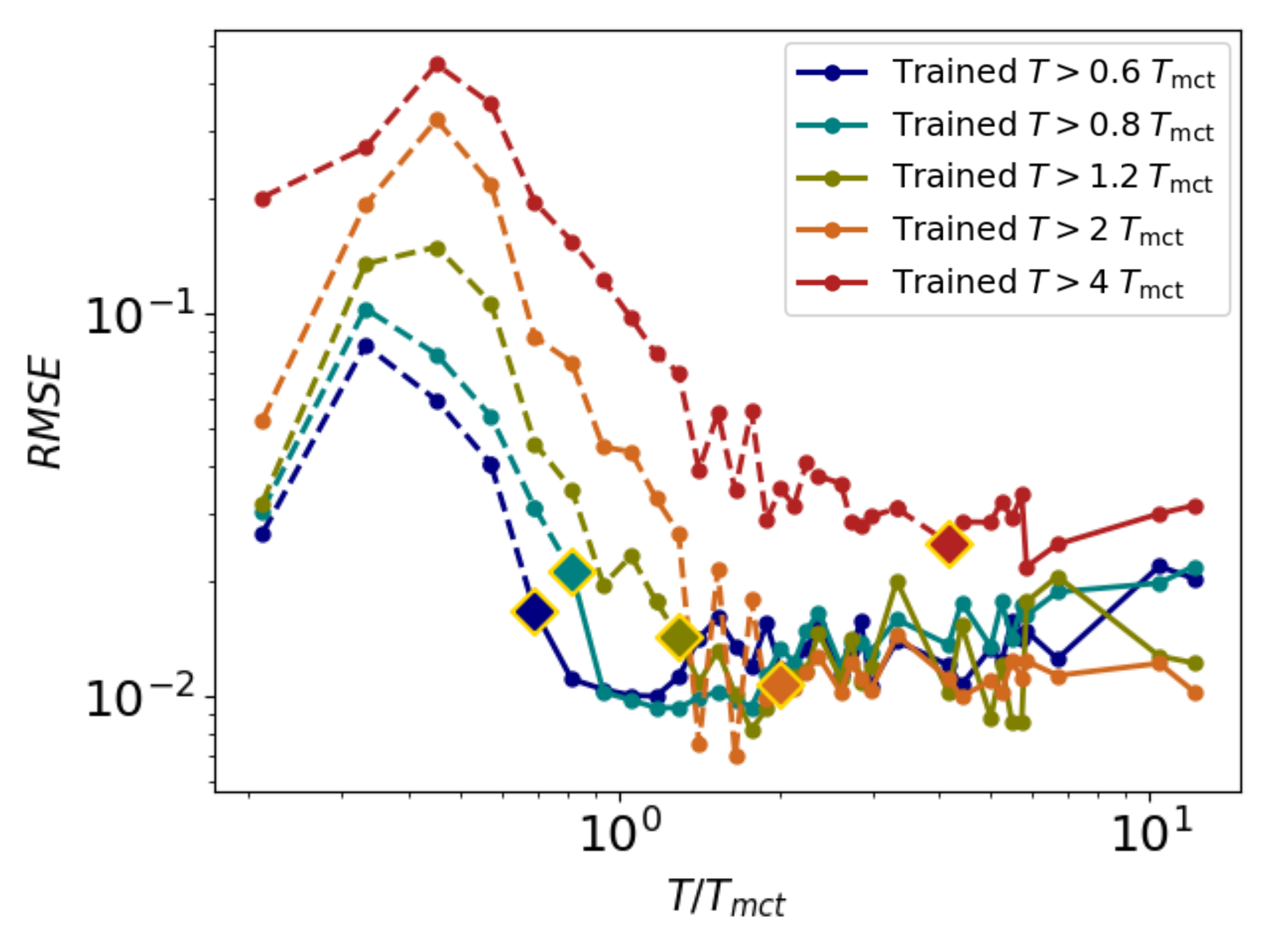}
\caption{Temperature transferability of the MLP. For each curve the model is trained using only temperatures larger than a certain value (indicated by the respective rhombus symbols). The solid lines represent the regions in which the model has been trained, while it has never seen data from the dashed line region.
}
\label{fig:Ttransf-rmse}
\end{figure}
\begin{figure}
\centering
\includegraphics[width=\columnwidth]{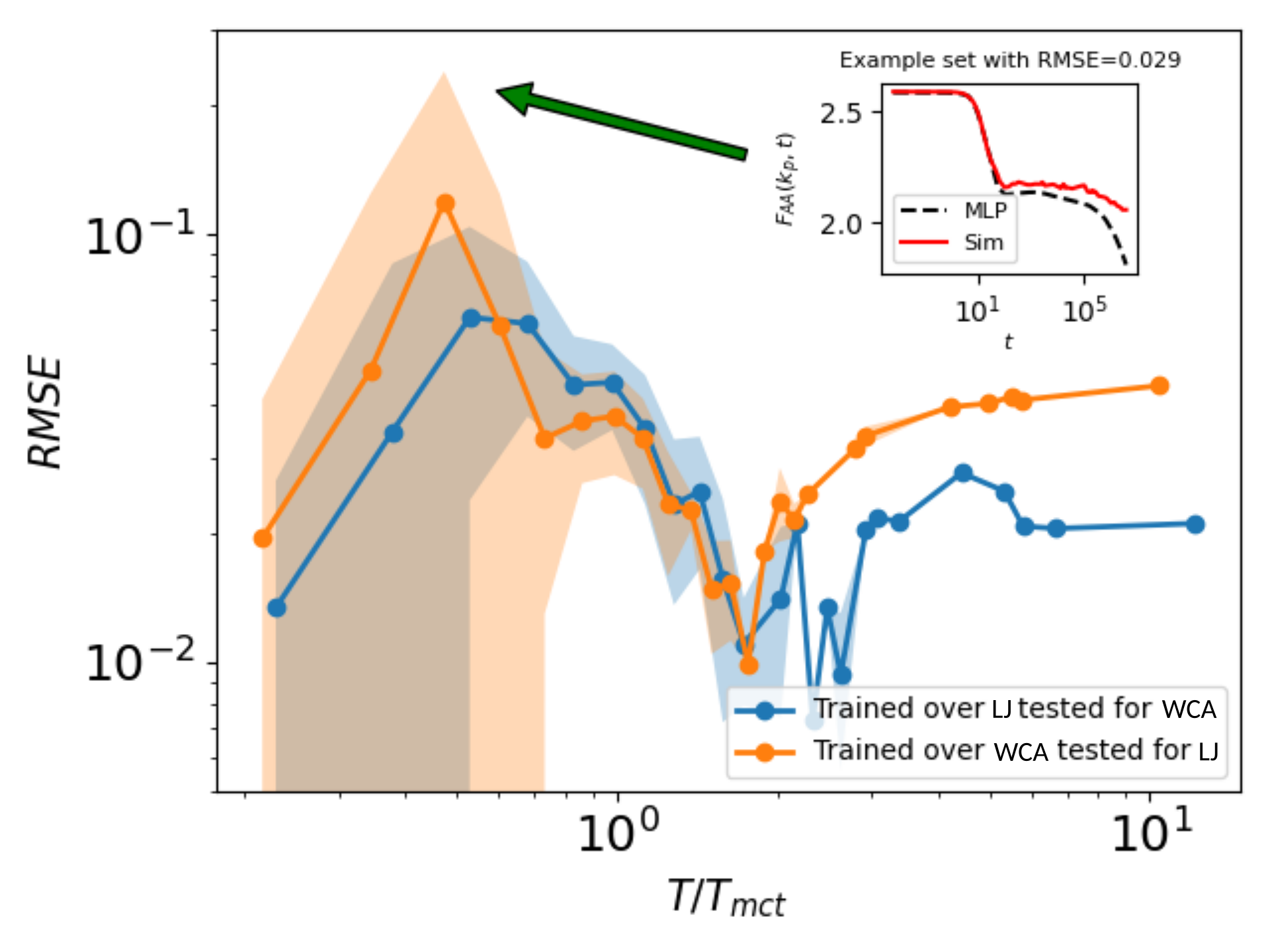}
\caption{Performance of the MLP when tested for a different interaction potential. The RMSE for the blue(orange) curve are the error in predicting the dynamics of LJ(WCA) simulations, while the model is trained only using WCA(LJ) data. 
The inset quantifies the RMSE error showing the discrepancy between prediction and ground truth.
}
\label{fig:modeltransf-rmse}
\end{figure}
Next, we report the RMSE (Eq.~\ref{eq:rmse}) corresponding to the temperature transferability (Fig.~\ref{fig:Ttransf-rmse}) and the model transferability (Fig.~\ref{fig:modeltransf-rmse}). The RMSE confirms the same conclusion we draw in the main manuscript: temperature/model transferability stops working approaching the experimental glass transitions. The RMSE also evidences the paradoxical result that very low temperature data are predicted more precisely as we can see from the drop in the RMSE for the leftmost points. As we explain in the main manuscript this is an artifact caused by limiting our observation at $t<10^7$, which means that at very low temperature we do not see $F$ leaving the plateau, so $F$ is an easier curve to reproduce for the MLP.

\begin{figure}
\centering
\includegraphics[width=\columnwidth]{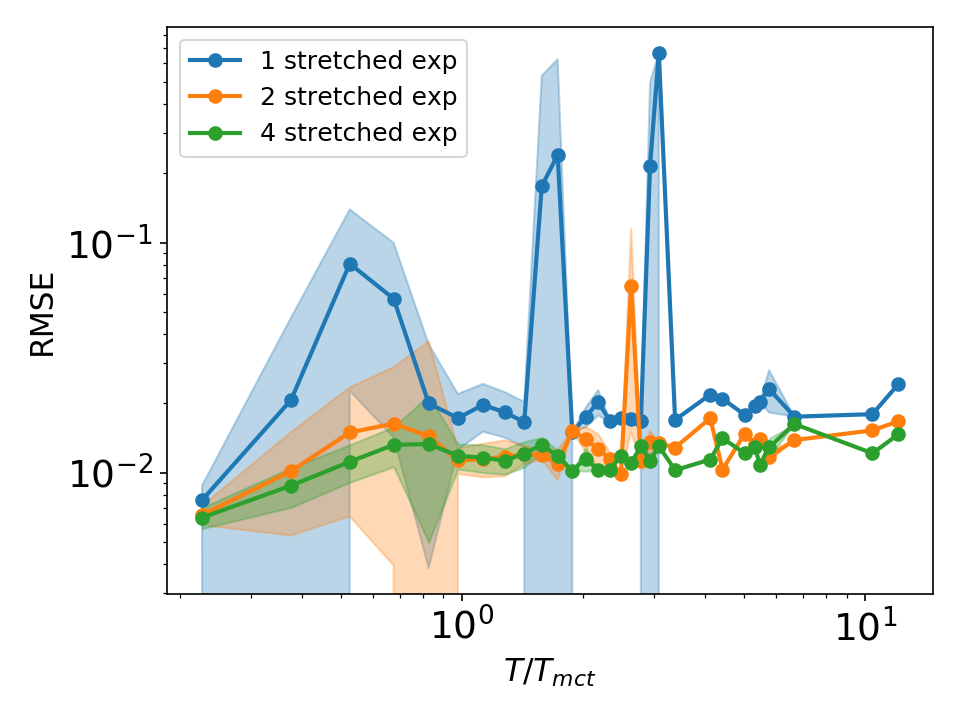}
\caption{Performance of the evolutionary strategy in finding the memory function that correctly predicts the dynamics. We parameterize the memory following Eq.~\ref{eq:strexp}, while the different curves correspond to different values of $N_{exp}$. We show that $N_{exp}=2$ already achieves good performances, so we conclude that $27$ parameters are enough to represent all the components of the memory function at fixed $k$.}
\label{fig:es-rmse}
\end{figure}
Finally we report the RMSE (Eq.~\ref{eq:rmse}) measure for the evolutionary strategy. In Fig.~\ref{fig:es-rmse}, we see that also the RMSE confirms that in order to achieve stable results, we need to represent the memory as at least 2 stretched exponentials, supporting the conclusion of the main manuscript. 

\subsection*{Evolutionary strategy targeting MCT}
\begin{figure}
    \centering
    \includegraphics[width=\columnwidth]{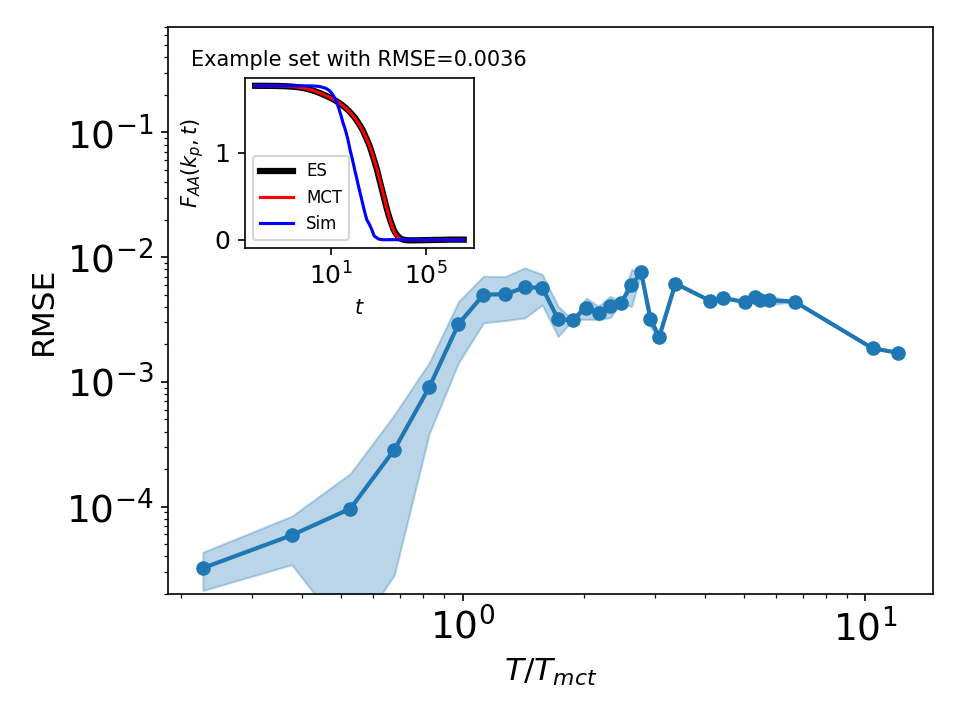}
    \caption{RMSE of the  same evolutionary strategy discussed in the main manuscript, but this time targeting the solution of the MCT equation $F_{mct}$. In the inset we see that the $F^{AA}(k_p,t)$ predicted by the ES converges towards MCT rather than the corresponding simulation.}
    \label{fig:es-mct}
\end{figure}
\begin{figure}
    \centering
    \includegraphics[width=\columnwidth]{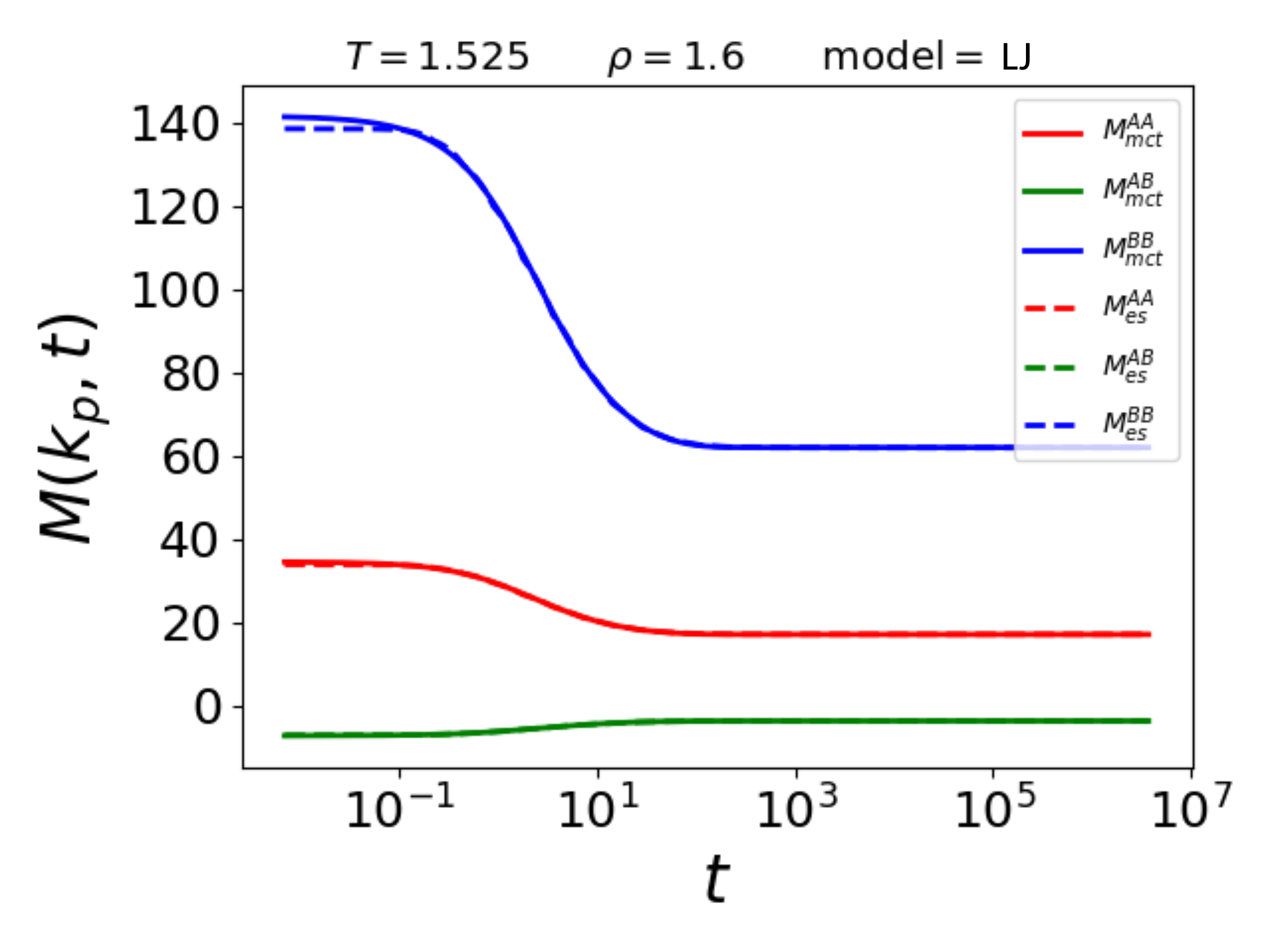}
    \caption{The three components of the memory function calculated by the ES reproduce almost perfectly the target $M_{mct}$.}
    \label{fig:m-mct}
\end{figure}
We justify here the legitimacy of our evolutionary strategy approach to the determination of the memory function $M$. 
The reason why this justification is needed is that machine-learning solutions are only evaluated by their loss function, while they could lose all the constraints that a physical function should have.
To show that our ES is \textit{physical}  we repeated the same analysis as sec.~\ref{sec:es}, but we targeted instead the dynamics generated by MCT, i.e. the intermediate scatting function $F$ that is the solution of the MCT equation.
We report in Fig.~\ref{fig:es-mct} the RMSE proving that the ES has converged to the target and it has learned $F_{mct}$. In Fig.~\ref{fig:m-mct} we show that the ES has effectively converged to $M_{mct}$, which is the exact memory that produced the dynamics $F_{mct}$.
In the main manuscript we target the real simulated dynamics where $M$ is unknown (which is also the reason why we developed this approach), but this convergence towards the MCT solutions constitutes an indication of the physicality of our evolutionary approach.

\begin{figure}
    \centering
    \includegraphics[width=\columnwidth]{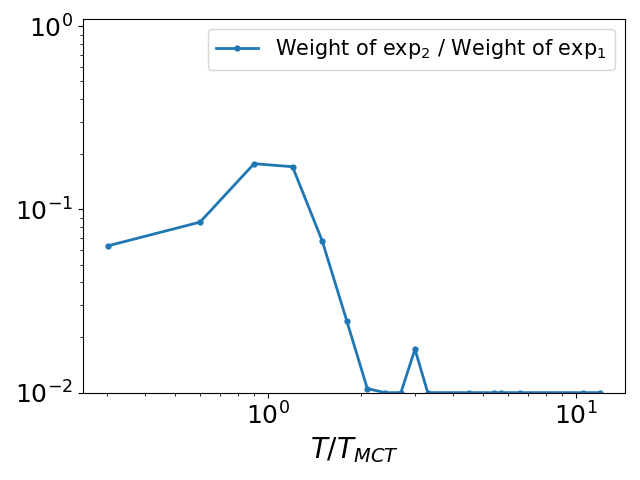}
    \caption{Comparison of the prefactor of the two stretched exponentials used to model the memory function in the main manuscript (Fig.~\ref{fig:es-r2}). We see that their ratio reaches $10\%$ only when the system is approaching the glass transition at $T\sim T_{MCT}$.}
    \label{fig:2strexp}
\end{figure}
\subsection*{Relative importance of the stretched exponentials}
We have seen in the main manuscript that the memory function can be parameterized as two stretched exponentials. From a theoretical level, this inspired us to propose a double stretched-exponential schematic MCT, reported in Eq.~\ref{eq:mctclosure} (main manuscript).
The two exponentials represent two different relaxation channels that allow $F$ to decrease over time. In this section of the supplementary we show that the second channel (i.e. the second exponential of Eq.~\ref{eq:strexp}) becomes relevant only approaching the glass transition. We see in fact in Fig.~\ref{fig:2strexp} that the prefactor of the second exponential reaches $10\%$ the value of the first, only when the system is approaching the glass transition. The values in Fig.~\ref{fig:2strexp} correspond to the convergence of the ES, reported in Fig.~\ref{fig:es-r2}.
The growth of the importance of the second exponential near $T_\mathrm{mct}$ is consistent with the fact that a simple exponential closure can only model mildly supercooled liquids~\cite{Gotze1992}.
We also know that supercooled liquids close to their glass transition show the emergence of activated relaxation and facilitation~\cite{guislein22,scalliet22} that could be associated to the second stretched exponential.
Overall, even if its microscopic nature is not made clear, the evolutionary strategy learns from the observed dynamics that approaching the glass transition the memory needs to develop a second stretched exponential.

\end{document}